# Data assimilation of fuel moisture in WRF-SFIRE

Martin Vejmelka[A,B,D], Adam K. Kochanski[C], and Jan Mandel[A]

[A]Department of Mathematics and Statistics, University of Colorado Denver,
1250 14[th] Street, Suite 600, Denver, CO, 80202
[B]Institute of Computer Science, Academy of Sciences of the Czech Republic, Prague, Czech Republic
[C]Department of Atmospheric Sciences, University of Utah, Salt Lake City, UT
[D] Corresponding author. E-mail: vejmelkam@gmail.com

**Abstract:** Fuel moisture is a major influence on the behavior of wildland fires and an important underlying factor in fire risk. We present a method to assimilate spatially sparse fuel moisture observations from remote automatic weather stations (RAWS) into the moisture model in WRF-SFIRE. WRF-SFIRE is a coupled atmospheric and fire behavior model which simulates the evolution of fuel moisture in idealized fuel species based on atmospheric state. The proposed method uses a modified trend surface model to estimate the fuel moisture field and its uncertainty based on currently available observations. At each grid point of WRF-SFIRE, this information is combined with the model forecast using a nonlinear Kalman filter, leading to an updated estimate of fuel moisture. We demonstrate the effectiveness of the method with tests in two real-world situations: a region in Southern California, where two large Santa Ana fires occurred recently, and on a domain enclosing Colorado.

**Additional Keywords:**

Introduction

The WRF-SFIRE model (Mandel *et al.* 2011) couples an established model of the atmosphere (WRF) (Skamarock *et al.* 2008), together with a model simulating fire behavior (SFIRE). Both components are connected via physical feedbacks — weather conditions enter the fire model and the emitted heat and vapor fluxes enter the weather model and directly perturb the state of the atmosphere in the vicinity of the fire. WRF-SFIRE has evolved from CAWFE (Clark *et al.* 2004). Similar models include MesoNH-ForeFire (Filippi *et al.* 2011). Recently, the WRF-SFIRE code has been extended by a fuel moisture model and coupled with the emissions model in WRF-Chem (Kochanski *et al.* 2012; Mandel *et al.* 2012). The current code and documentation are available from OpenWFM.org. A version from 2010 is distributed with the WRF release as WRF-Fire (Coen *et al.* 2012; OpenWFM 2012).

The behavior of fire is highly sensitive to fuel moisture content. Evaporation of moisture in the fire consumes heat, which cannot then contribute to fire propagation. With increasing fuel moisture content, the spread rate decreases, and eventually, at the extinction moisture level, the fire does not propagate at all (Pyne et al. 1996). The fuel moisture content depends on vegetation properties and on atmospheric conditions.

This paper reports on an effort to use fuel moisture observations supplied by remote automatic weather stations (RAWS) to adjust the state of the fuel moisture model in WRF-SFIRE.

**Methods**

*The moisture model*

The fuel moisture model in WRF-SFIRE (Kochanski *et al.* 2012; Mandel *et al.* 2012) simulates $N_k$ idealized, homogeneous fuel species. Such fuel species are commonly referred to by their drying/wetting time lag $T_k$ as 1-hour, 10-hour, and 100-hour fuel (Pyne *et al.* 1996). The moisture content in each idealized fuel species is simulated on a coarse grid, while the actual fuel used in the fire propagation is a mixture of these species on a much finer grid, where the fire simulation takes place. At each point of the coarse grid, the moisture content of each fuel species is simulated independently by a first order differential equation with time lag $T_k$. The solution of the differential equation approaches asymptotically an equilibrium fuel moisture content. The equilibrium depends on atmospheric conditions (temperature, relative humidity, pressure) and on whether the current fuel moisture approaches the equilibrium from above (drying) or from below (wetting). If the fuel moisture is between the drying and the wetting equilibria, it does not change. The effect of rain is modeled by the same type of time-lag equation, with the time lag value dependent on the rain intensity.

Denote the fuel moisture content of the $k$-th idealized fuel species with time lag $T_k$ by $m_k$, stored as a dimensionless proportion of kg of water per kg of wood. The fuel moisture model is described mathematically by the ordinary differential equation

$$\frac{d}{dt} m_k = \begin{cases} \dfrac{S - m_k(t)}{T_r}\left(1 - \exp\left(\dfrac{r(t) - r_0}{r_k}\right)\right) & \text{if } r(t) > r_0 \text{ (soaking in rain)}, \\ \dfrac{E_d(t) - m_k(t)}{T_k} & \text{if } r(t) \leq r_0, m_k(t) > E_d(t) \text{ (drying)}, \\ \dfrac{E_w(t) - m_k(t)}{T_k} & \text{if } r(t) \leq r_0, m_k(t) < E_w(t) \text{ (wetting)}, \\ 0 & \text{otherwise}, \end{cases} \quad (0.1)$$

where $E_d(t)$ is the drying equilibrium, $E_w(t)$ is the wetting equilibrium, $S$ is the rain saturation level, $r_0$ is the threshold rain intensity, $r(t)$ is the current rain intensity, $r_k$ is the saturation rain intensity, $T_k$ is the drying/wetting time lag, and $T_r$ is the asymptotic soaking time lag in very high-intensity rain. The fuel coefficients $T_k$, $r_k$, and $r_0$ can be specified for each idealized fuel class by the user. By default, the equilibria $E_d(t)$ and $E_w(t)$ are computed from WRF atmospheric variables at the Earth surface following the Canadian fire danger rating model (Van Wagner and Pickett 1985). In particular, $E_d(t) - E_w(t) > 0$ is constant. For the rain regime, by default, the equilibrium is taken as $S = 2.5$ and the coefficients $T_r$, $r_0$, and $r_k$ were identified to match the behavior of the fuel soaking in rain in Van Wagner and Pickett (1985). The differential equation is solved by a numerical method exact for any length of the time step, for constant coefficients. This is important because fuel moisture modeling may be done on a much larger time scale (hours) than fire behavior modeling (seconds), cf. Kochanski *et al.* (2012) for further details.

The present method assimilates observations into current fuel moisture level $m_k(t)$ and the equilibria $E_d(t)$, $E_w(t)$, and $S$. We shall postpone the reasons for not assimilating the time lags to the discussion. Since the equilibria are computed from external quantities, a standard solution is to extend the state of the model to also contain perturbations of the equilibria. Adding the perturbations to the model (0.1), we get an extended dynamical system for the variables $m_k$, $\Delta E$, and $\Delta S$,

$$\frac{d}{dt} m_k = \begin{cases} \dfrac{S + \Delta S(t) - m_k(t)}{T_r} \left(1 - \exp\left(\dfrac{r(t) - r_0}{r_k}\right)\right) & \text{if } r(t) > r_0 \\ \dfrac{E_d(t) + \Delta E(t) - m_k(t)}{T_k} & \text{if } r(t) \leq r_0,\ m_k(t) > E_d(t) + \Delta E(t) \\ \dfrac{E_w(t) + \Delta E(t) - m_k(t)}{T_k} & \text{if } r(t) \leq r_0,\ m_k(t) < E_w(t) + \Delta E(t) \quad (0.2) \\ 0 & \text{otherwise,} \end{cases}$$

$$\frac{d}{dt} \Delta E = 0,$$

$$\frac{d}{dt} \Delta S = 0.$$

We write the discretization of the extended model (0.2) as

$$\mathbf{m}(t_{i+1}) = f(\mathbf{m}(t_i), E_d(t_i), E_w(t_i), r(t_i)), \qquad (0.3)$$

where the extended fuel moisture model state is

$$\mathbf{m}(t_i) = (m_1(t_i), m_2(t_i), \ldots, m_{N_k}(t_i), \Delta E(t_i), \Delta S(t_i)). \qquad (0.4)$$

The introduction of the shared assimilated parameters $\Delta E$ and $\Delta S$ transforms the isolated equations for each fuel species into a coupled system. The coupling provides a natural pathway for propagating observation-based state updates from the observed fuel species to the unobserved species within the dynamical model. The effect of data assimilation of 10-hr fuel moisture observations on other types of fuel (1-hr, 100-hr) will be investigated elsewhere, whereas in this paper we focus on the effect of data assimilation on the model state of the 10-hr fuel.

*Extended Kalman Filter*

We use the Extended Kalman Filter (EKF) (Simon 2010, §13.2.3) to assimilate independently the fuel moisture field at each grid point as estimated by the trend surface model presented in the next section. At each time step, the EKF combines the extended moisture model state $\mathbf{m}(t)$ given by (0.4) with the information given by the moisture data and its variance.

The EKF tracks the evolution of the state mean and covariance, assuming that the initial state was normally distributed, $\mathbf{m}(t_0) \sim \mathrm{N}\,(\mathbf{m_0}, P_0)$. As the initial state $\mathbf{m_0}$, we use the equilibrium moisture at $t_0$, which is computed from WRF variables, and estimated covariance $P_0$. Subsequently, we model the evolution of the fuel moisture using the function $f$ from (0.3) with the Jacobian

$$J_f(t_i) = \nabla f(\mathbf{m}(t_{i-1}), E_d(t_{i-1}), E_w(t_{i-1}), r(t_{i-1})).$$

The observation operator is $g(\mathbf{m}(t_i)) = m_l(t_i)$, and its Jacobian is

$$J_g(t_i) = \nabla g(\mathbf{m}(t_i)) = (0, \ldots, 1, \ldots 0),$$

where 1 is at the position of the observed fuel species in the extended state vector. The EKF first predicts the new mean and variance of the model state as

$$\widehat{\mathbf{m}}(t_i) = f(\mathbf{m}(t_{i-1}), E_d(t_{i-1}), E_w(t_{i-1}), r(t_{i-1}))$$
$$\hat{P}(t_i) = J_f(t_{i-1}) P(t_{i-1}) J_f^T(t_{i-1}) + Q,$$

where $Q$ is the process noise covariance. If no observations of fuel moisture are available at $t_i$, then the predicted mean and covariance become the mean and covariance of the state distribution in the next time step,

$$\mathbf{m}(t_i) = \widehat{\mathbf{m}}(t_i), \quad P(t_i) = \hat{P}(t_i).$$

If, however, observations $\mathbf{d}(t_i)$, with error covariance $S_i$, are available, then the mean and covariance of the state are updated according to the formulas

$$K(t_i) = \hat{P}(t_i) J_g^T(t_i) \left( J_g(t_i) \hat{P}(t_i) J_g^T(t_i) + S_i \right)^{-1}$$
$$\mathbf{m}(t_i) = \widehat{\mathbf{m}}(t_i) - K(t_i)(g(\widehat{\mathbf{m}}(t_i)) - \mathbf{d}(t_i))$$
$$P(t_i) = \left( I - K J_g(t_i) \right) \hat{P}(t_i).$$

*Transporting of observations to grid points*

We use a variant of the trend surface modeling approach (Schabenberger and Gotway 2005, §5.3.1) to transport observed information across space from the RAWS locations to each grid point. We prefer this method to a full universal kriging approach and argue our viewpoint in the discussion section. The assumed form of fuel moisture observation $\mathbf{Z}(s)$ at location $s$ is

$$\mathbf{Z}(s) = \beta_1 \mathbf{X}_1(s) + \mathrm{L} + \beta_k \mathbf{X}_k(s) + e(s) = \mathbf{x}(s)\boldsymbol{\beta} + e(s),$$

where the fields $\mathbf{X}_j$, called covariates, are known at every location $s$, $\boldsymbol{\beta}_j$ are unknown regression coefficients, the error $e(s)$ is independent at each grid point and $\mathbf{x}(s) = [\mathbf{X}_1(s), \mathbf{X}_2(s), \ldots, \mathbf{X}_k(s)]$ is a row vector of covariates at location $s$. The error $e(s)$ is assumed to have zero mean and consist of an independent observation error with variance $\gamma^2(s)$, assumed to be known, and an unobservable microscale variability with variance $\sigma^2$, which is the same at every location $s$ (Cressie 1993). We write the observation model in compact matrix form,

$$\mathbf{Z} = \mathbf{X}\boldsymbol{\beta} + e, \, e \sim \mathrm{N}(0, \Sigma), \quad \Sigma = \Gamma + \sigma^2 I. \qquad (0.5)$$

where $\Gamma = \text{diag}(\gamma^2(s))$.

The coefficients $\boldsymbol{\beta}_k$ and the microscale variability variance $\sigma^2$ are estimated from the data. Given the microscale variance $\sigma^2$, observations $\mathbf{Z}_s$, and covariates $\mathbf{X}_s$ at the same locations $\mathbf{s} = (s_1, s_2, \ldots, s_n)$, the regression coefficients $\hat{\boldsymbol{\beta}}$ are determined from weighed least squares as

$$\hat{\boldsymbol{\beta}} = (\mathbf{X}_s^T \boldsymbol{\Sigma}_s^{-1} \mathbf{X}_s)^{-1} \mathbf{X}_s^T \boldsymbol{\Sigma}_s^{-1} \mathbf{Z}_s, \qquad (0.6)$$

where $\boldsymbol{\Sigma}_s$ is the submatrix of the covariance matrix corresponding to the locations of the observations. To estimate the microscale variability variance $\sigma^2$, we numerically solve the equation

$$\sum_{i=1}^{n} \frac{\hat{e}(s_i)^2}{\gamma^2(s_i) + \hat{\sigma}^2} = n - k \qquad (0.7)$$

for $\hat{\sigma}^2$, where $\hat{e}_i(s_i) = \hat{\mathbf{Z}}(s_i) - \mathbf{x}(s_i)\hat{\boldsymbol{\beta}}$ are the regression errors at location $s_i$. Both $\hat{\beta}$ and $\hat{\sigma}^2$ are found by an iterative method starting from $\hat{\sigma}^2 = 0$: In each iteration, the method first computes $\hat{\boldsymbol{\beta}}$ from (0.6) and then $\hat{\sigma}^2$ from (0.7). The observation $\mathbf{d}(t_i)$ injected into the EKF at the grid point $s$ is then $\mathbf{x}(s)\hat{\boldsymbol{\beta}}$, with the estimated variance

$$\hat{\sigma}^2 + \mathbf{x}(s)(\mathbf{X}(\mathbf{s})^T \boldsymbol{\Sigma}(\mathbf{s})^{-1} \mathbf{X}(\mathbf{s}))^{-1} \mathbf{x}(s)^T.$$

Mathematical derivation and properties of this estimator will be studied elsewhere.

We use up to $k = 8$ covariates $X_i$. Some of the covariates can be omitted by the user. The most important covariate is the current forecast of fuel moisture from the WRF-SFIRE moisture model. Additionally, three other variables are extracted from WRF: the temperature at 2 m, the surface pressure and the current rain intensity, all strongly affecting the fuel moisture equilibrium. These covariates capture the effect of local atmospheric state on the evolution of fuel moisture and thus can be expected to approximate the spatial structure of the fuel moisture field. The remaining four covariates are constant in time and model the effects of spatial and topographical features on fuel moisture. They are latitude, longitude, elevation of each grid point, and a constant vector. The latitude and longitude facilitate modeling of domain-scale spatial linear trends. The elevation accounts for the effect of terrain profile on fuel moisture, and the constant covariate allows for the adjustment of a mean difference between the model and the observation stations. The strategy of using multiple covariates in the model aims to capture as much of the spatial structure in the deterministic component of the model as possible since the random component of a trend surface model does not account for any spatial dependence structure.

**Results**

We have performed two case studies. For the first one, the region of Southern California has been selected, where two massive Santa Ana fires (Witch Creek and Guejito) burned 56,796 ha in October 2007, leading to $18M in damage and two fatalities (Keeley et al. 2009). For the second test case, a period of six consecutive days was selected with the domain covering the State of Colorado, with no fire. The two case studies use the WRF-SFIRE model based on WRF 3.3 and WRF 3.4 configurations, respectively. No data assimilation has been performed in the atmospheric model, which was initialized and driven by the North American Regional Reanalysis (Mesinger et al. 2006).

WRF output data was stored in 10 minute intervals, which was also the time step for the moisture model in the presented test cases. In the data assimilation step, all observations within 30 min of the current time were gathered and processed. Larger time windows have a smoothing character, while smaller time windows produce more time-resolved snapshots of the fuel moisture field but introduce larger variance into the data assimilation step as fewer observations enter the assimilation step at each time point. The 30-min assimilation window was selected as a balance between the two requirements based on multiple simulations with different time windows.

The RAWS which supplied fuel moisture observations have 10-hr fuel stick sensors and their observations and metadata were obtained from the MesoWest[1] website. The fuel moisture observations are provided as the number of grams of water in 100 g of pine wood. Before assimilation, these are rescaled to a dimensionless value in the range 0 to 1, in order to match the representation of the fuel moisture in the model.

As data on the variance of fuel moisture observations from RAWS was not available, a constant variance of $\gamma^2(s) = 0.01^2$ was attributed to each observation, which corresponds to a standard deviation of 0.01. In future, it is expected that more accurate data differentiating the quality of measurements from RAWS will be available. When available, this information can then be directly used by the current method.

The assimilation algorithm uses the full set of eight covariates in the trend surface model in both simulations. During the simulation, $N_k = 3$ idealized fuel species were modeled: 1-hr, 10-hr and 100-hr fuel. We focus on reporting the results obtained for the 10-hr fuel.

---

[1] http://mesowest.utah.edu/

The Kalman filter was initialized at each grid point with the equilibrium fuel moisture at the time of the start of the simulation with $P_0 = 0.01I$, indicating a high uncertainty in the initial state. The process noise covariance for the ten minute time step was chosen as

$$Q = \text{diag}(1\times10^{-4}, 5\times10^{-5}, 1\times10^{-5}, 5\times10^{-5}, 5\times10^{-5})$$

*Southern California*

The San Diego area of Southern California has been selected to test the applicability of the algorithm, as a region experiencing frequent severe wildfires associated with Santa Ana events. During these periods, strong Santa Ana winds bring very hot and dry air from the Nevada desert, rapidly reducing the fuel moisture and increasing fire danger. This test case presents an example of a realistic application of the fuel moisture data assimilation algorithm. The objective is to spin-up the fuel moisture content prior to a fire simulation and increase the accuracy of fire moisture and fire spread forecasts. With this in mind, the spin-up time should be short, so as not to impact the total spin-up time of the model prior to ignition, as simulating fire behavior is computationally demanding.

The simulation was performed for a period from 7/21/2007 12:00 UTC to 7/24/2007 12:00 UTC in a 4-domain configuration. The outer domain (d01), responsible for resolving the large-scale flow responsible for generating the Santa Ana winds, had a resolution of 36km and covered a region of 4320×3072 km. A set of 3 finer domains has been nested within this domain, in order to gradually provide more detailed representation of the terrain and meteorological conditions in the area of interest. The nested domains (d02, d03 and d04) had resolutions of 12 km, 4 km and 1.33 km respectively (see Fig. 1). The 3D atmospheric state was resolved on a vertically-stretched grid, with 37 levels of gradually decreasing vertical resolution — from 20 to 500 m. The domain nesting used in this study is shown in Fig. 1.

Hourly fuel moisture observations were available from ten RAWS located in the domain d04 as shown in Fig. 2. Only the RAWS station Palomar Mountain (PAMC1) had missing values in the simulated time period – 17 observations.

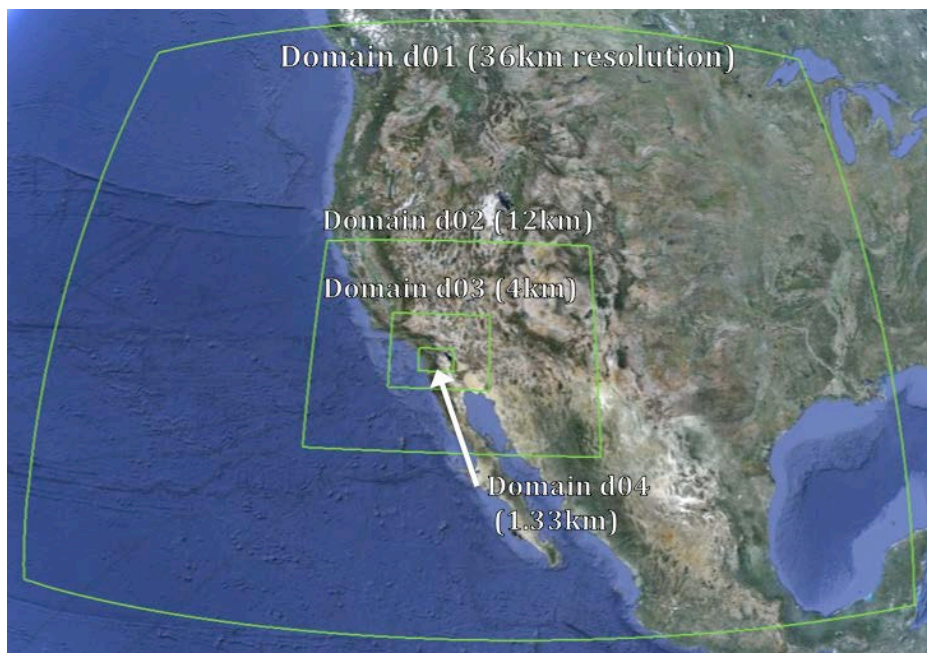

**Fig. 1:** The nested domain configuration for the Southern California simulation.

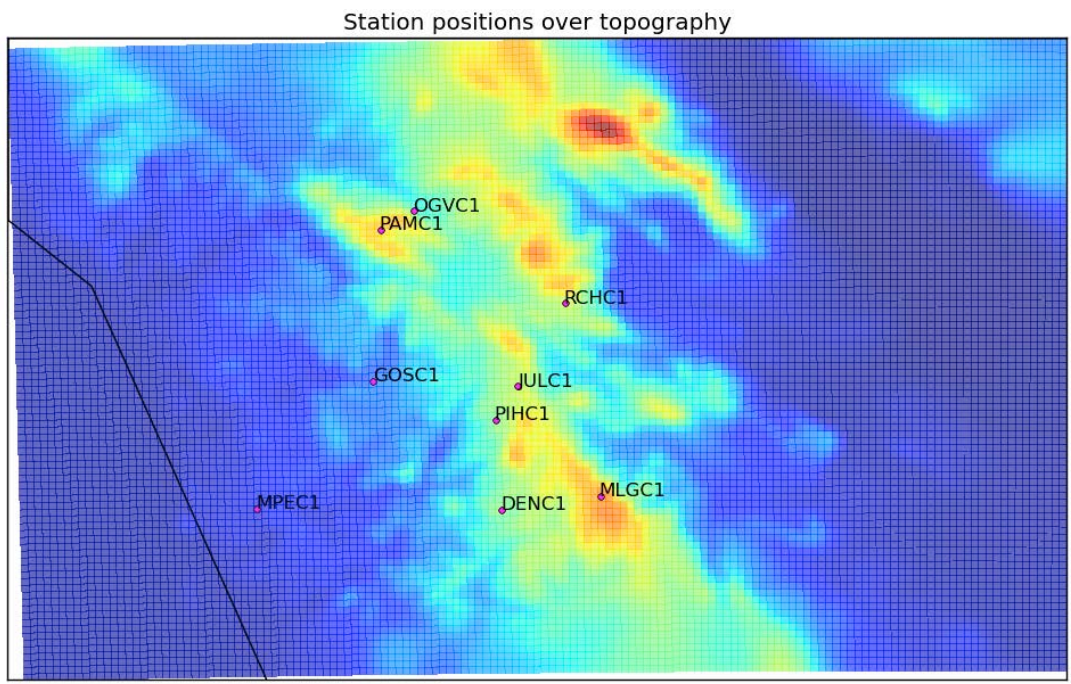

**Fig. 2:** Positions of stations supplying fuel moisture observations in the domain d04 overlaid on topography.

The estimates of microscale variability variance are in the bottom panel of Fig 3. The estimates of the microscale variability are initially high as WRF itself spins up and the spatial structure of the fuel moisture field in WRF is initialized from equilibrium conditions. After about 4 h, the microscale variability variance stabilizes and does not increase above 0.005. The mean square differences of the assimilated model state and station observations and of the model with no assimilation and the station observations are shown in Fig. 3, top panel. At the time the microscale variability variance stabilizes, the mean squared difference between the observations also reaches its stable value.

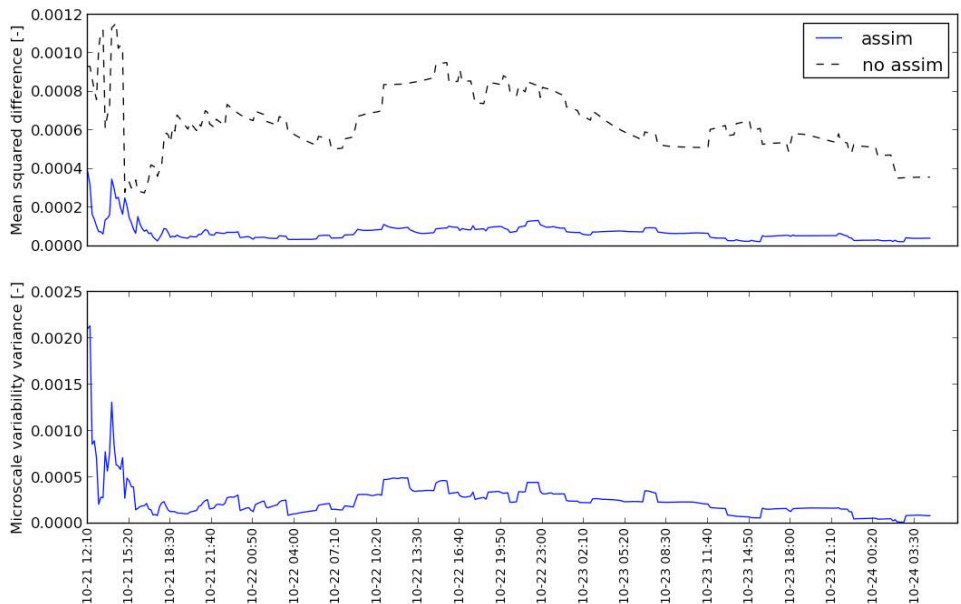

**Fig. 3:** The mean square differences between non-assimilated and assimilated fuel moisture state estimates and RAWS observations (top) and microscale variability variance estimates as a function of time (bottom) for the Southern California test.

An example 10-hr fuel moisture distribution map obtained towards the end of the simulated timespan is shown in Fig. 4.
　We summarize that in this study the spin-up time was short (4 h) and it will not impact the total spin-up time of the fire simulation prior to fire ignition. The data assimilation method was able to reduce the difference between observations and model forecasts substantially. While in this study, the number of covariates ($k = 8$) was close to the number of RAWS stations observed, there were no apparent issues of numerical stability.

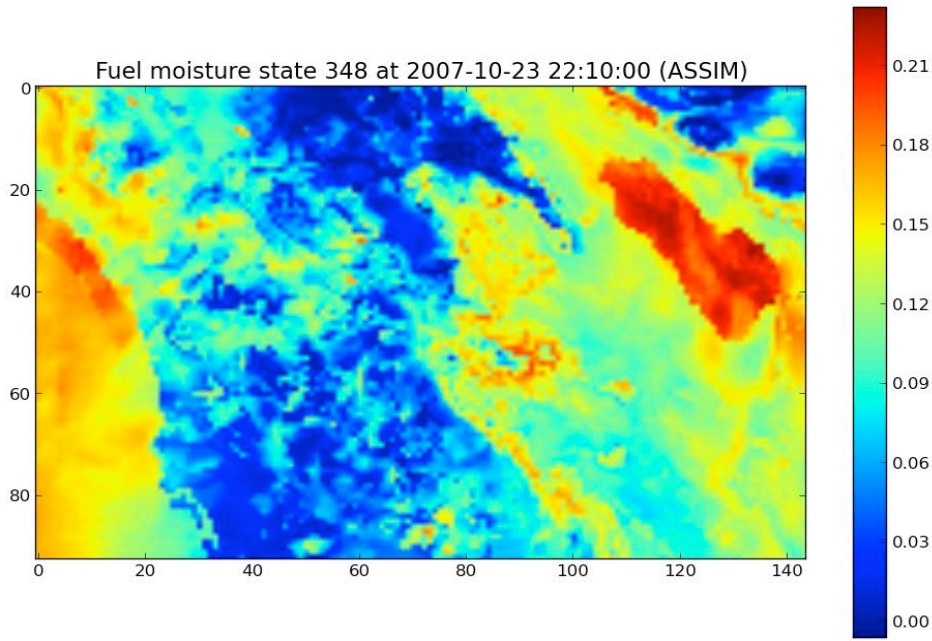

**Fig. 4:** A fuel moisture map of the 10-hr idealized fuel species obtained during the Southern California simulation run. In the bottom left part of the image, the coastline is clearly visible and in the top right corner of the image, the Salton Sea is prominent. Time is in UTC.

*Colorado*

The Colorado experiment was selected as a large-scale scenario representing the use case where the presented algorithm will be applied to estimate moisture content of dead fuel operationally for the purposes of fire risk estimation. Colorado presents a difficult challenge for any data assimilation algorithm, as an area experiencing complex weather patterns. The interactions between the synoptic flow and Colorado's mountainous landscape lead to local weather conditions with strong spatial variations. The dramatic changes in elevation and insolation translate into significant near-ground moisture and temperature variations across relatively small spatial distances, which are often too fine to be captured by a relatively coarse grid of a mesoscale model. The sharp gradients in meteorological variables due to high spatial variability of the meteorological fields and topography may also numerically destabilize the model and deteriorate the quality of weather forecasts. Colorado also often experiences severe droughts that lead to high fire danger and severe fires (like the Fourmile Canyon Fire (2010) and Waldo Canyon Fire (2012)), which makes it a suitable site for testing and deployment of fuel moisture assimilation algorithms.

    The WRF simulation was run from 6/1/2013 00:00 UTC to 6/6/2013 00:00 UTC in a single domain configuration covering Colorado with a 2 km grid with 264 x 200 nodes. Observations

were obtained from a total of 42 RAWS in the Colorado area positioned as shown in Fig. 5. This set of RAWS was selected out of a total of 52 RAWS listed as active in the Colorado area on the MesoWest website after removing stations yielding suspicious data (sensors yielding only one value indicating possible damage) and stations that supplied no observations in the given timespan. The 10-hr fuel moisture observations from the RAWS were supplied hourly.

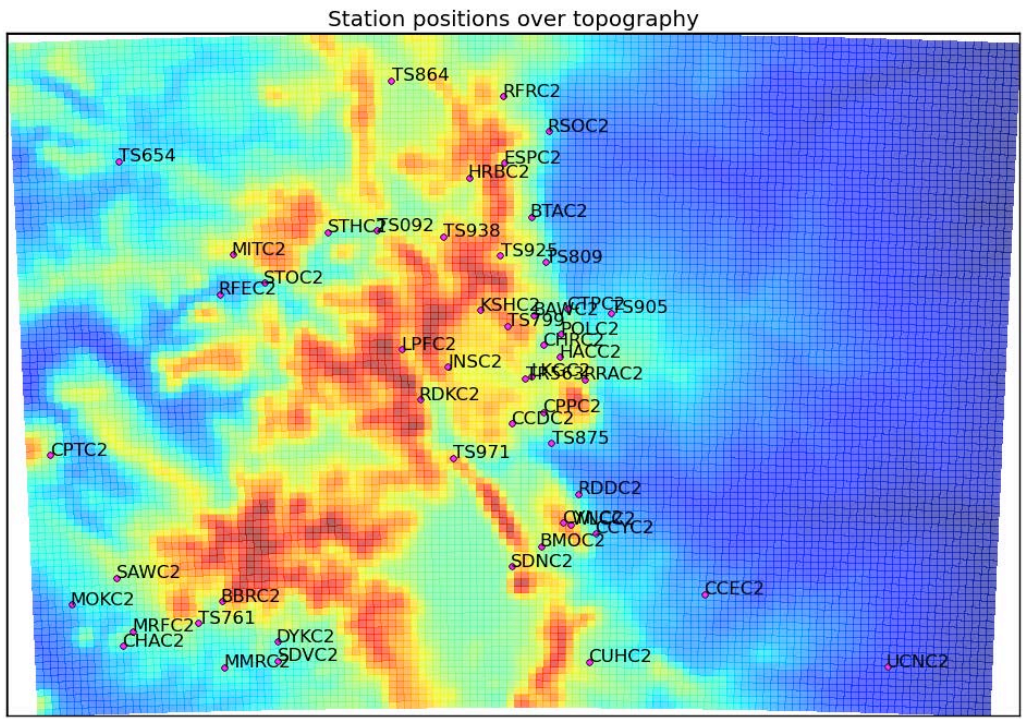

**Fig. 5:** Positions of stations supplying fuel moisture observations overlaid on Colorado topography.

The estimates of microscale variability are shown in Fig. 6, bottom panel. There are three spikes in the microscale variability coinciding with precipitation events captured by WRF-SFIRE. At 2 km spatial resolution, the simulation may not resolve these structures sufficiently, in addition to being possibly inaccurate without data assimilation in the atmospheric state. The trend surface model uses covariates obtained from the WRF model and it is thus unable to immediately capture the local changes in fuel moisture caused by precipitation. The estimates of microscale variability variance are thus increased in these episodes, which increases the uncertainty of the fuel moisture values from the trend surface model entering the extended Kalman filters at each grid location.

The mean square differences of the assimilated model state and station observations are shown in Fig. 6, top panel. As in the Southern California test case, it is clear that the mean square differences are substantially reduced even during precipitation events. An example fuel moisture map for 10-hr fuel is shown in Fig. 7.

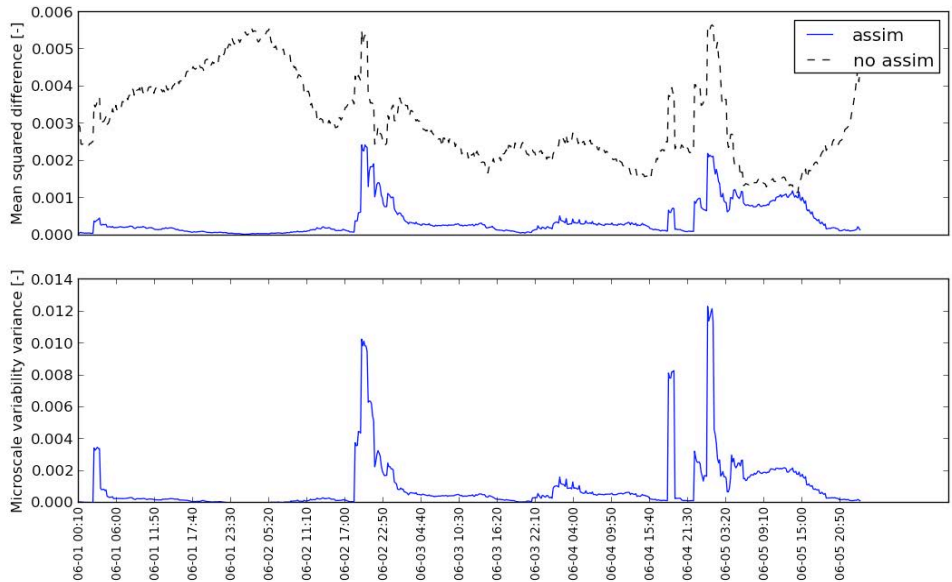

**Fig. 6:** The mean square differences between non-assimilated and assimilated fuel moisture state estimates and RAWS observations (top) and microscale variability variance estimates as a function of time (bottom) for the Colorado test case.

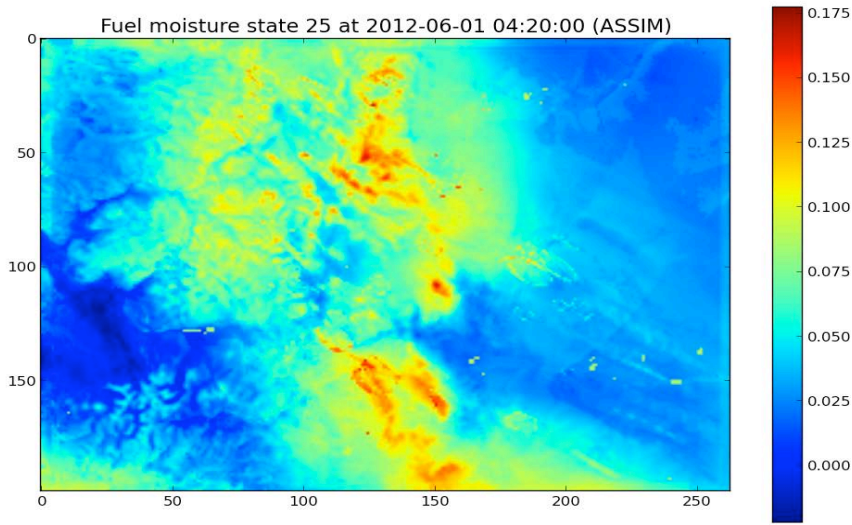

**Fig. 7:** An example fuel moisture map of the 10-hr idealized fuel species obtained toward the beginning of the Colorado simulation run. The time is in UTC.

This simulation shows that although complex atmospheric phenomena may increase the uncertainty of the fuel moisture estimates temporarily, the data assimilation system does not exhibit computational instability in these instances and is able to recover quickly.

**Discussion**

*Assimilating the time lags of idealized fuels*

The assimilation of time lags for the idealized fuels is entirely possible but not useful in the context of the aims of the data assimilation method. Fuel moisture from the idealized fuel species is transformed into estimates of fuel moisture in one of thirteen different vegetation classes by weighted averaging, which requires 1-hour, 10-hour and 100-hour idealized fuel species as input. By assimilating time lags, we would in effect model different idealized fuel species than required by the subsequent users of the results. Nevertheless, during development of the method simulation runs were conducted with assimilation of time lags and it was found that the effect of data assimilation on time lags was negligible. Assimilation of fuel time lags is thus neither desired nor effective.

*Trend surface modeling and universal kriging*

The objective of the proposed method is its integration in an operational fuel moisture assimilation mechanism. Strong emphasis on the stability and predictability of the numerical algorithms is thus important in addition to minimal user intervention requirements. In complex terrain, a complicated model of covariance, perhaps including multiple terrain characteristics, would be necessary to exploit any potential spatial relationships in the observed data. An examination of variograms of fuel moisture observations in the Front Range region of Colorado has not uncovered a convincing distance-related structure. The question of the ultimate efficacy and stability of a complex covariance model remain. Finally, universal kriging is typically used in much smaller or much larger domains, at scales where assumptions on smoothness of the terrain and ambient conditions facilitate the construction of distance-based models of covariance. At the mesoscale range of the current simulations, non-stationarity induced by weather phenomena and terrain properties makes use of universal kriging methods challenging in the least.

*Future developments*

In future research, use of data assimilation methods adjusting the atmospheric state is expected to significantly improve the function of the presented assimilation mechanism, especially during the appearance of transient complex weather patterns.

The effects of data assimilation on other modeled idealized fuel species (1-hr, 100-hr) will be investigated. These fuel species are only indirectly affected by the 10-hr observations due to coupling in the extended model state and no direct observations of the fuel moisture of these species are available to the authors at this time that would facilitate a performance analysis of the assimilation method.

**Conclusion**

In this paper, we have presented a method for assimilation of remote automatic weather station 10-hr fuel moisture observations into the fuel moisture model in WRF-SFIRE. The method was constructed for the primary purposes of improving fire behavior modeling in on-demand fire modeling scenarios and for operational fire risk estimation. We have demonstrated on two real-world examples that the proposed method adjusts the model forecasts towards the station observations, is able to function in complex topography and recover from transient disturbances, such as precipitation events.

**Acknowledgments**

This research was partially supported by the National Science Foundation (NSF) grants AGS-0835579 and DMS-1216481, National Aeronautics and Space Administration (NASA) grants NNX12AQ85G and NNX13AH9G, and the Grant Agency of the Czech Republic grant 13-34856S. The authors would like to thank the Center for Computational Mathematics University of Colorado Denver for the use of the Colibri cluster, which was supported by NSF award CNS-0958354. This work partially utilized the Janus supercomputer, supported by the NSF grant CNS-0821794, the University of Colorado Boulder, University of Colorado Denver, and National Center for Atmospheric Research.

**References**

Clark TL, Coen J, Latham D (2004) Description of a coupled atmosphere-fire model. *International Journal of Wildland Fire*, **13**, 49–64.
Coen J, Cameron JM, Michalakes J, Patton E, Riggan P, Yedinak K (2012) WRF-Fire: Coupled weather-wildland fire modeling with the Weather Research and Forecasting model. *Journal of Applied Meteorology and Climatology*, **52**, 16–38.
Cressie, NAC (1993) 'Statistics for Spatial Data.' (John Wiley & Sons Inc.: New York)
Filippi JB, Bosseur F, Pialat X, Santoni P, Strada S, Mari C (2011) Simulation of coupled fire/atmosphere interaction with the MesoNH-ForeFire models. *Journal of Combustion*, **2011**, 540390.
Keeley JE, Safford H, Fotheringham CJ, Franklin J, Moritz M (2009) The 2007 Southern California Wildfires: Lessons in Complexity. *Journal of Forestry*, **107**, 287–296.
Kochanski AK, Beezley JD, Mandel J, Kim M (2012) WRF fire simulation coupled with a fuel moisture model and smoke transport by WRF-Chem. 13th WRF Users' Workshop, National Center for Atmospheric Research, (Boulder, CO), arXiv:1208.1059.
Mandel J, Beezley JD, Kochanski AK (2011) Coupled atmosphere-wildland fire modeling with WRF 3.3 and SFIRE 2011. *Geoscientific Model Development*, **4**, 591–610.
Mandel J , Beezley JD, Kochanski AK, Kondratenko VY, Kim M (2012) Assimilation of perimeter data and coupling with fuel moisture in a wildland fire – atmosphere DDDAS. *Procedia Computer Science*, **9**, 1100–1109.

Mesinger F, DiMego G, Kalnay E, Mitchell K, Shafran PC, Ebisuzaki W, Jovic D, Woollen J, Rogers E, Berbery EH, Ek MB, Fan Y, Grumbine R, Higgins W, Hong L, Ying L, Manikin G, Parrish D, and Shi W (2006) North American regional reanalysis. *Bulletin of the American Meteorological Society*, **87**, 343–360.

OpenWFM, 2012: Fire code in WRF release. Open Wildland Fire Modeling e-Community, http://www.openwfm.org/wiki/Fire_code_in_WRF_release, accessed November 2012.

Pyne S, Andrews PL, Laven RD (1996) 'Introduction to Wildland Fire.' (Wiley: New York)

Schabenberger O, Gotway CA (2005) Statistical Methods for Spatial Data Analysis. Chapman and Hall/CRC, Boca Raton.

Simon, D (2010) 'Optimal State Estimation: Kalman, H Infinity, and Nonlinear Approaches.' (Wiley: Hoboken).

Skamarock WC, Klemp JB, Dudhia J, Gill DO, Barker DM, Duda MG, Huang XY, Wang W, Powers JG (2008) A description of the Advanced Research WRF version 3. NCAR Technical Note 475, http://www.mmm.ucar.edu/wrf/users/docs/arw_v3.pdf, retrieved December 2011.

Van Wagner CE, Pickett TL (1985) 'Equations and FORTRAN program for the Canadian forest fire weather index system.' Canadian Forestry Service, Forestry Technical Report 33.